\documentclass[amsmath,amssymb,aps,prx,twocolumn]{revtex4-2}

\usepackage{graphicx}
\usepackage{dcolumn}
\usepackage{bm}
\usepackage[mathlines]{lineno}
\usepackage{color}

\usepackage[dvipdfmx,
colorlinks=true,
urlcolor=blue,
citecolor=blue,
linkcolor=blue,
hyperfootnotes=false]{hyperref}

\begin{document}

\title{Statistical Mechanics Approach to the Holographic Renormalization Group
:\\ Bethe Lattice Ising Model and p-adic AdS/CFT}

\author{Kouichi Okunishi${}^{1}$}\email{okunishi@phys.sc.niigata-u.ac.jp}
\author{Tadashi Takayanagi${}^{2,3,4}$}
\affiliation{${}^1$Department of Physics, Niigata University,  Niigata 950-2181, Japan}
\affiliation{${}^2$Center for Gravitational Physics and Quantum Information,Yukawa Institute for Theoretical Physics, Kyoto University,Kyoto 606-8502,Japan}
\affiliation{${}^3$Inamori Research Institute for Science,Kyoto 600-8411 Japan}
\affiliation{${}^4$Kavli Institute for the Physics and Mathematics of the Universe(WPI), University of Tokyo, Chiba 277-8582,Japan}

\date{\today}

\begin{abstract}
The Bethe lattice Ising model ---a classical model of statistical mechanics for the phase transition--- provides a novel and intuitive understanding of the prototypical relationship between tensor networks and Anti-de Sitter (AdS)/conformal field theory (CFT) correspondence. 
After analytically formulating a holographic renormalization group for the Bethe lattice model, we demonstrate the underlying mechanism and the exact scaling dimensions for the power-law decay of boundary spin correlations by introducing the relation between the lattice network and an effective Poincare metric on a unit disk. 
We compare the Bethe lattice model in the high-temperature region with a scalar field in AdS$_2$, and then discuss its more direct connection to the p-adic AdS/CFT.
In addition, we find that the phase transition in the interior induces a crossover behavior of boundary spin correlations, depending on the depth of the corresponding correlation path.
\end{abstract}

\maketitle


\section{ Introduction \label{tag_sec1}}

In accordance with the development of Anti-de Sitter (AdS)/conformal field theory (CFT) correspondence\cite{Maldacena1998,Gubser1998,Witten1998,Aharony2000}, the holography principle\cite{tHooft1993,Susskind1994,Busso2002} has been one of the most fascinating concepts relevant to various fields of theoretical physics.
In particular,  the holographic entanglement entropy defined by minimal surface in the AdS space-time\cite{RT_PRL2006,RT_JHEP2006} revealed the importance of quantum information involved in the space-time geometry, like area law of the entanglement entropy\cite{Eisert_RMP2010}, and created novel cross-disciplinary research frontiers of quantum gravity, quantum information, condensed matter physics, and cosmology. 
Among various developments, a prominent example associated with quantum many-body physics is the multi-scale entanglement renormalization ansatz (MERA)\cite{MERA2007}, known as a scale-invariant tensor network, providing a new pathway to bridge between the holography principle and the quantum many-body physics in condensed matter/statistical physics.\cite{Swingle2012}
In fact, inspired by the MERA, novel tensor-network-based mechanisms have been proposed for geometrical controlling of entanglements in quantum field theories, like continuous MERA\cite{cMERA2013,Nozaki2012} and efficient optimization of path integrals\cite{Caputa2017}.

Historically, the tensor network developed as an efficient theoretical tool for solving quantum/classical many-body systems in condensed matter physics, based on the network of integrated tensors representing the propagation of quantum entanglement.\cite{JPSJ2022}
After cooperating with quantum information physics in the 21st century, the tensor network has been established as a key theoretical building block in modern quantum physics\cite{Orus2019}, stimulating the interdisciplinary developments associated with the holography principle.
Indeed, such fascinating ideas as holographic error correction code\cite{Happy} and holographic random tensor networks\cite{Hyden2016} have been proposed so far, suggesting that the tensor network framework has an intrinsic connection to the holography principle.\cite{Jahn2019}
In addition, the tensor-network format provides essential insight for designing classical simulations of quantum circuits.\cite{Zhou2020,GordonBell}
However, a thorough understanding of the relationship between AdS/CFT and tensor networks beyond their network shape remains one of the most challenging problems since tensor network states comprising the MERA-type network are usually designed as a variational state for the ground-state problem on the CFT side.

The aim of this work is to provide a new perspective on the relationship between tensor networks and the holographic renormalization group (RG)\cite{deBoer2000} from the statistical-mechanics side.
In the landscape of theoretical physics, statistical mechanics often provides a simple and intuitive view behind nontrivial and essential concepts.
In particular, the analysis of 2nd order phase transitions and critical phenomena played a crucial role in establishing the concept of RG, which is the most fundamental foundation of modern theoretical physics.\cite{WilsonRG,Kadanoff2014}
Also, the formal correspondence between the quantum and classical systems based on the Suzuki-Trotter decomposition has been extensively utilized in lattice simulations of quantum many-body systems.\cite{Suzuki1976,Trotter1959}
In the context of the tensor network, the density matrix renormalization group and related tensor network algorithms for quantum systems\cite{White1992,Schollwock2005} share the unified framework of matrix product states with the corner transfer matrix\cite{Baxter1968}  for classical statistical mechanics systems.\cite{JPSJ2022}
Moreover, the theoretical structure of RG-based entanglement controlling behind MERA became more visible in the tensor-network renormalization group for 2D classical spin models.\cite{TNR2015,Evenbly2015}.
How can we describe the holographic structure of the AdS space-time in the language of a lattice spin model?
Answering this question from the statistical-mechanics viewpoint is expected to be a crucial step toward filling in a missing piece between the tensor network and AdS/CFT.

Recently, holographic properties of hyperbolic lattice models, which are a possible lattice counterpart of AdS$_2$, have attracted much attention.\cite{Shima2006,UedaK2007,Krcmar2008,Lee2016,Asaduzzaman2020,Asaduzzaman2022}
In particular, numerical simulations for hyperbolic lattice spin models suggest that the bulk spin correlation function decays exponentially\cite{Iharagi2010,Gendiar2012}, while intriguing power-law decay of correlation functions was observed for boundary spins\cite{Asaduzzaman2022}.
However, a lack of analytical understanding based on holographic RG prevents us from exploring the further connection of hyperbolic lattice models to AdS/CFT.

\begin{figure}[tb]
\begin{center}
\includegraphics[width=7cm]{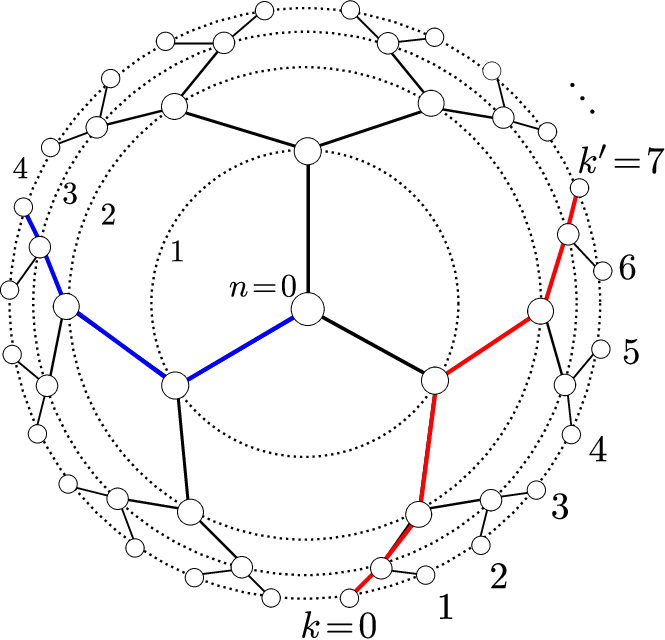}~~
\end{center}
\caption{The Ising model on a Cayley tree lattice with $N=4$, where the index $n$ for dotted circles represents the generation from the center spin.
The radius of the circles is defined by Eq. (\ref{eq_radius}) with $z_n$, which asymptotically corresponds to the radial coordinate of the  Poincare coordinate.
The distance between $k=0$ and $k'=2^3 -1 $ is defined by counting the number of nodes along the outer edge of the Cayley tree lattice, while the red line represents the one-dimensional path connecting the edge spins.}
\label{fig1}
\end{figure}

In this paper, we analytically formulate a holographic RG for the Bethe lattice Ising model by introducing unit-disk representation to the Bethe lattice network (See Fig. \ref{fig1}), which can be asymptotically described by the Poincare metric.
The Bethe lattice model was originally introduced as a spin model on a self-similar tree network attributed to the Bethe approximation for the corresponding regular-lattice model, and its bulk phase transition is characterized by the mean-field universality class.\cite{Bethe1935,Domb1960,BaxterBook}
Nevertheless, we find that this classic spin system, which is also viewed not only as a limit of the hyperbolic lattice model\cite{Mosseri1982,Daniska2016} but also as the simplest example of the tree tensor network, involves minimal physics of the AdS$_2$; 
The power-law behavior with nontrivial exponents is remarkably observed for the boundary spins rather than the bulk region, which enables us to investigate the holographic RG structure analytically. 
By comparing a scalar field in AdS$_2$, we discuss that the Bethe-lattice Ising model can be regarded  as a prototype lattice model of a toy version of AdS$_2$/CFT$_1$.
We also demonstrate its more direct relationship to the p-adic AdS/CFT\cite{Gubser2017,Heydeman:2016ldy,Bhattacharyya:2016hbx,Bhattacharyya:2017aly,Gubser:2017tsi,Hung:2019zsk}, which provides an intriguing connection of the statistical lattice model to the AdS/CFT.
In addition, the bulk phase transition associated with the Bethe approximation induces a crossover behavior of the boundary spin correlations, depending on how the correlation path penetrates the bulk regime.

The remainder of this paper is organized as follows.
In Sec. 2, we reformulate the recursive method of solution for the Bethe lattice Ising model as a holographic RG.
In Sec. 3, we explain the relation of the holographic RG for the Bethe lattice model to the power-law decay of boundary spin correlation functions by introducing a unit disc coordinate.
In Sec. 4, then, we discuss the correspondence between the Bethe lattice Ising model and a scalar field in the AdS$_2$.
Moreover, we elucidate its more direct relationship to the p-adic AdS$_2$.
In Sec. 5, we show a crossover behavior of the boundary correlation functions in the low-temperature region and discuss its interpretation based on the tree tensor network.
Sec. 6 is devoted to the summary and discussions.

\section{The Bethe lattice Ising model toward the holographic RG}

Let us consider a Bethe lattice Ising model of $N$th generation with a coordination number $q(>2)$.
As an example, we illustrate a Cayley tree lattice Ising model($q=3$) of $N=4$ in Fig. \ref{fig1}.
The partition function of the Bethe lattice Ising model is defined as 
\begin{align}
\mathcal{Z}_N = \sum_{\{\sigma\}} \exp\left(K \sum_{\langle i,j \rangle}\sigma_i \sigma_j  + h_b \sum_{i=\mathrm{edge}} \sigma_i\right)
\label{eq_betheZ}
\end{align}
where $\sum_{\langle i,j \rangle}$ denotes the sum with respect to the neighboring pairs on the Bethe lattice as in Fig. \ref{fig1} and $\sum_{i=\mathrm{edge}}$ means the sum of spins on the outer-edge boundary nodes.
Also, $K$ and $h_b$ respectively represent an inverse temperature and a boundary magnetic field for the outer-edge spins.
Here, note that no bulk magnetic field is applied in Eq. (\ref{eq_betheZ}).   
Although the basic property of Eq. (\ref{eq_betheZ}) was intensively studied in 70s\cite{Domb1960,Eggarter1974,Mullar-Hartmann1974,Morita1975,BaxterBook, Hu1998}, we reveal the intrinsic connection between the Bethe-lattice Ising model and the holography by reformulating its solution from the viewpoint of the holographic RG.

A particular point of the Bethe lattice is that although it has a tree network structure with spreading branches, Eq. (\ref{eq_betheZ}) can be mapped into an effective one-dimensional (1D) Ising model, for which one can take a partial sum with respect to the spins recursively from the outer edge. 
This is because any paths connecting $\sigma_0$ and the boundary spins are equivalent, and the $q-1$ number of the descendent branches for any node in the Bethe lattice network also gives the same contribution to the node.
Thus, it is sufficient to pick up a 1D path connecting the center spin $\sigma_0$ and a boundary spin $\sigma_N$ in the outer edge circle, as illustrated as a blue line in Fig. \ref{fig1}.
Then, it is useful to follow Baxter's recursive representation of $\mathcal{Z}_N$\cite{BaxterBook}.
Explicitly we have
\begin{align}
\mathcal{Z}_N = \sum_{\sigma_0} [g_N(\sigma_0)]^q  \, ,
\label{eq_zbaxter}
\end{align}
with 
\begin{align}
g_N(\sigma_0) & = \sum_{\sigma_1}e^{K\sigma_0\sigma_1} [g_{N-1}(\sigma_1)]^{p} \nonumber \\
&= \sum_{\sigma_1}e^{K\sigma_0\sigma_1 } \left[\sum_{\sigma_2} 
e^{K\sigma_1\sigma_2 }[g_{N-2}(\sigma_2)]^{p}\right]^{p} \nonumber \\
&= \cdots \, ,
\label{eq_rbaxter}
\end{align}
which provides the recursive definition of $g_n(\sigma_{N-n})$ based on the partial sum of the larger generation spins with the boundary condition $g_0(\sigma_N)=e^{h_b\sigma_N}$.
Here we note that 
\begin{align}
p\equiv q-1
\end{align}
on $g_n(\sigma_{N-n})$ for $n\ne 0$ is attributed to the number of descendent branches.

In Eq. (\ref{eq_rbaxter}), $[g_{N-n}(\sigma_{n})]^{p}$ represents the realization probability of $\sigma_{n}=\pm 1$.
We can then reformulate $[g_{N-n}(\sigma_{n})]^{p} \propto \exp(\tilde{h}_n\sigma_n)$ with introducing an effective magnetic field $\tilde{h}_n$. 
This implies that the effect of the descendent spins is recursively renormalized into the effective magnetic field $\tilde{h}_{n-1}$ for the parent generation spin $\sigma_{n-1}$.
Explicitly, we can construct the relation
\begin{align}
C_{n-1} e^{\tilde{h}_{n-1}\sigma_{n-1} } =  \left [ \sum_{\sigma_n} e^{K\sigma_{n-1}\sigma_{n} + \tilde{h}_n \sigma_n} \right]^{p},
\label{eq_effmag}
\end{align}
starting from $\tilde{h}_N\equiv h_b$.
Equivalently, we have  
\begin{align}
\tilde{h}_{n-1} = f(\tilde{h}_n),
\label{eq_betheRG}
\end{align}
with 
\begin{align}
f(x) \equiv \frac{p}{2}\log\left(\frac{e^{K+x}+e^{-K-x}}{e^{K-x}+ e^{-K+x}} \right).
\end{align}
Note that 
\begin{align}
C_{n-1}=  \left[4\cosh(K+\tilde{h}_n)\cosh(K-\tilde{h}_n)\right]^{\frac{p}{2}}
\end{align}
is an overall coefficient independent of the spin configuration.
For Eq. (\ref{eq_betheRG}), it is worthwhile pointing out that $\tilde{h}_n$ affects only $\tilde{h}_{n-1}$ while the renromalization of $K$ never occurs.
Using Eq. (\ref{eq_betheRG}) recursively, we then obtain the partition function as
\begin{align}
\mathcal{Z}_N = \left( \prod_{n=0}^{N-1} C_n \right) \sum_{\sigma_0} \exp\left( \frac{(p-1)\tilde{h}_0}{p}   \sigma_0 \right)
\label{eq_bethe_prt}
\end{align}
which is reduced to a single spin problem in the effective magnetic field $\tilde{h}_0$ corresponding to Eq. (\ref{eq_zbaxter}).
Then the magnetization at the center spin is straightforwardly obtained as $M_0 = \tanh(q \tilde{h}_0 /(q-1))$.

\begin{center}
\begin{figure*}[tb]
\begin{center}
\includegraphics[width=13cm]{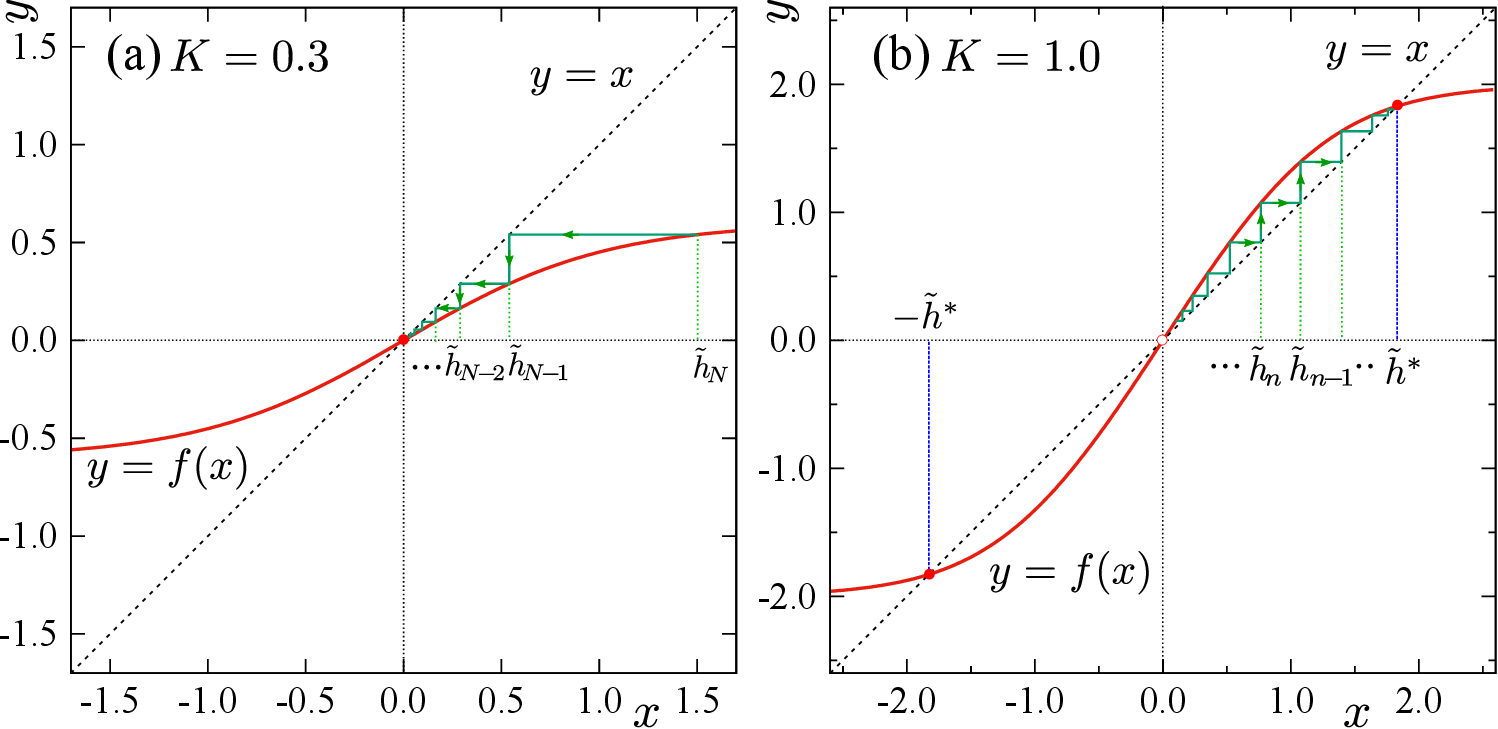}~~
\end{center}
\caption{The recursive relation for $\tilde{h}_n$ for the Cayley tree Ising model ($q=3$), where solid red curves represent $y=f(x)$ and dotted lines are $y=x$.
(a) $K=0.3$($<K_B=0.463\cdots$), where $\tilde{h}_n$ is absorved into the fixed point $\tilde{h}_*=0$, and (b) $K=1.0$, where $\tilde{h}_n$ flows into a nontrivial fixed point of $\pm \tilde{h}_*$ corresponding to the spin order in the low-temperature phase.
}
\label{fig2}
\end{figure*}
\end{center}

For analyzing the flow property of $\tilde{h}_n$,  a promising way is to use a graph technique as depicted in Fig. \ref{fig2}.
We can draw two possible types of the $\tilde{h}_n$ flow diagrams depending on $K$, which reflect the phase transition in the sense of the Bethe approximation.
Our idea is that this $\tilde{h}_n$ flow can be interpreted as a holographic RG flow defined in an effective hyperbolic space.
As in Fig. \ref{fig2}, then, $\tilde{h}_n$ approaches to the trivial fixed point $\tilde{h}_*=0$  for $K<K_B$, while for $K>K_B$ the flow is absorbed into nontrivial fixed points $\pm \tilde{h}_*$, which is determined by the fixed point equation
\begin{align}
\tilde{h}_* = f(\tilde{h}_*) \, .
\label{eq_bethe_fixed}
\end{align}
Of course, this fixed point relation is nothing but the self-consistent equation corresponding to the Bethe approximation for the Ising model with a coordination number $q$.
The trivial solution $\tilde{h}_*= 0$ in Fig. \ref{fig2} (a) describes the disordered state in the high-temperature region.
In the low-temperature region, the nontrivial fixed point $\tilde{h}_*$  provides the magnetic order in the deep interior as in Fig. \ref{fig2} (b),
The transition point $K_B$ is explicitly obtained as
\begin{align}
K_B = (T_B)^{-1} & \equiv  \frac{1}{2}\log \left( \frac{q}{q-2} \right) \nonumber \\
 &= \arctan \left( \frac{1}{p} \right) \, ,
\label{eq_Tb}
\end{align}
where $y=f(x)$ and $y=x$ are in contact with each other at $x=0$.

Here it should be remarked that the $K_B$ does not mean the bulk phase transition for the entire Bethe lattice system, although the deep interior exhibits distinct fixed point structures across $K_B$.
This is because the dominant number of spins are in the vicinity of the outer edge boundary, and thus the deep interior basically does not cause any singularity of the bulk free energy as a function of $K$, in consistency with the fact that the total system can be described by the effective 1D chain.
In this paper, we do not come into this problem, which was precisely discussed in Refs. \cite{Eggarter1974,Mullar-Hartmann1974}.

In the context of the holography,  we focus on the behavior of $\tilde{h}_n$ near the outer edge boundary and then discuss how the effect of the boundary field $h_b$ is enhanced/reduced toward the interior.
First, let us consider the perturbative regime of $h_b \ll 1$ for the bulk fixed point $\tilde{h}_*=0$.
By linearizing $f(x)$ around $x=0$, we obtain the $n$-dependence of $\tilde{h}_n$ as
\begin{align}
\tilde{h}_{n} \simeq h_b \lambda^{N-n}
\label{eq_bare_hn}
\end{align}
with 
\begin{align}
\lambda= p\tanh K\, .
\label{eq_lambda}
\end{align}
If $K<K_B$, $\lambda < 1$ and thus $\tilde{h}_n \to 0$ as $n$ decreases from $N$.
This behavior corresponds to the case where $h_b$ is irrelevant in the sense of the RG flow from the boundary to the interior.
\footnote{In the sense of AdS/CFT, $h_b$ is a relevant field for a boundary CFT.} 
While for $K>K_B$, $\tilde{h}_n$ enhances toward $\tilde{h}_n \to \tilde{h}_* (\ne 0)$ as in Fig. \ref{fig1}(b), and thus $h_b$ is relevant.

Another important quantity for the argument based on holography is the correlation function of spins located on the outer edge boundary.
As an example, we consider a spin pair connected by the red-line path in Fig. \ref{fig1} and introduce an additional index $k$ indicating the position of boundary spins along the outer edge circle, $\sigma_{Nk}$, for later convenience. 
If $h_b=0$, then, one can straightforwardly calculate the correlation function for arbitrary spin pairs as 
\begin{align}
\langle \sigma_{Nk} \sigma_{Nk'} \rangle =  (\tanh K )^{2d_{kk'}} \, ,
\label{eq_corr}
\end{align}
where $2d_{kk'}$ denotes the number of bonds along the lattice network path (correlation path) between $k$ and $k'$ nodes.\cite{Morita1975}
Using the generation index $n_{kk'}$ for the spin node at the middle of the correlation path\footnote{$n_{k,k'}$ corresponds to the smallest generation index along the correlation path}, we can explicitly write
\begin{align}
d_{kk'}= N-n_{kk'} \, .
\label{eq_def_d}
\end{align}
For example, $N=4$, $k=0$, $k'=7$, $n_{kk'}=1$ and $d_{kk'}=3$ for the red-line path in Fig. \ref{fig1}.
The result of Eq. (\ref{eq_corr}) with Eq. (\ref{eq_def_d}) implies that the correlation function exhibits exponential decay with respect to the distance along the correlation path in the unit of node index on the Bethe lattice, reflecting the 1D nature of the tree network structure.

\section{The Bethe lattice and Poincare coodinate}

\subsection{The Bethe lattice Ising model on a disk}

In the lattice Hamiltonian (\ref{eq_betheZ}), the interaction between any adjacent spins is homogeneous everywhere on the lattice. 
As in Fig. \ref{fig1}, on the other hand,  the Bethe lattice Ising model drawn on a disk illustrates that the radii of circles become inhomogeneous and the density of spins becomes dense as the generation index $n$ increases. 
In order to describe this situation of the Bethe lattice Ising model quantitatively, we introduce a disk-like coordinate with inhomogeneous radii.
More precisely, we define the radius of the circle with the generation index $n$  as 
\begin{align}
R_n = 1-z_n\, ,
\label{eq_radius}
\end{align}
with
\begin{align}
z_n \equiv  p^{-n} \, ,
\label{eq_def_z}
\end{align}
for which the interval between two circles of the generations $n$ and $n+1$ is scaled with $p^{-n}$ and the radius of the outer edge with $N \to \infty$ is normalized to be unity.
Also, $p(>1)$ corresponds to the number of branches except for the center cite, and $z_n$ represents the distance from the outer edge circle ($R_\infty=1$), which can be asymptotically related to the radial variable in the Poincare coordinate.
Meanwhile,  $z_0 = \frac{p-1}{p}$ at the center site of the Bethe lattice implies that the lattice space in the infrared (IR) limit is fixed to be of the order of unity.
Note that the circles in Fig. \ref{fig1} were illustrated with Eq. (\ref{eq_radius}).

Using the radial coordinate of Eq. (\ref{eq_def_z}),  then, we rewrite Eq. (\ref{eq_bare_hn}) to
\begin{align}
\tilde{h}_n \sim  \lambda^{-n} =z_n^{1- \Delta}
\label{eq_h_z}
\end{align}
with
\begin{align}
\Delta \equiv - \frac{\log(\tanh K)}{\log p}\, .
\label{eq_Delta}
\end{align}
In the disk representation, we can thus observe the power-raw behavior of $\tilde{h}_n$ with the scaling dimension $1-\Delta$ near the outer edge boundary, instead of the exponential decay of Eq. (\ref{eq_bare_hn}).

Similarly, we need to describe the spin correlation function along the outer edge boundary in association with the $z_n$ coordinate.
As mentioned before, a particular point on the Bethe lattice,  which is a tree-type tensor network, is that all descendent branches of a certain node site are equivalent.
This implies that the Bethe lattice model may not have an intrinsic metric defining the relative relation of spins in the circumference direction, which enabled us to obtain the effective 1D representation of the partition function $\mathcal{Z}_N$. 
{\it We therefore arrange the spin nodes having the same generation index $n$ uniformly on a circle with a constant radius};
As in Fig. \ref{fig1}, we take a usual polar coordinate with the radius $R_n$ and assign the index $k$ in the circumference direction. 
Then the spin node on the $n$th generation circle is specified by   
\begin{align}
x_k = \frac{2\pi R_n}{qp^{n-1}} k  
\simeq \frac{2\pi p  }{q} z_n  k \quad  \mathrm{for} \quad n \gg 1 \,   ,
\label{eq_def_x}
\end{align}
where the location of $k=0$ is assumed to be the leftmost branch in one of the biggest subtree networks for later convenience.
This definition of $x$ coordinate will turn out to be natural in discussing a connection to the Poincare coordinate near the outer edge boundary.

We turn to the analysis of the correlation function. 
Taking account of the tree network structure, we consider the spin pair attached at the leftmost branch and the rightmost branch in a sub-tree network, as indicated by the red path in Fig. \ref{fig1}.
Then, the number of spins between $k=0$ and $k'$ is $p^{d_{0k'}}$, while the total number of the spins along the outer edge circle is $qp^{N-1}$.
For $N \gg 1$, thus, the distance of the spin pair along the outer edge circle can be described as 
\begin{align}
x_{k'} - x_0 \simeq \frac{2 \pi p^{d_{0k'}}}{q p^{N-1}} \, ,
\label{eq_def_r}
\end{align}
Here, note that this definition may contain an ``error" due to the nature of the Bethe-lattice network;
For instance, $x_{k'} - x_0$ does not have a naive translational invariance along the circumference direction, as in the case of the p-Adic CFT.

\begin{figure}[tb]
\begin{center}
\includegraphics[width=6.5cm]{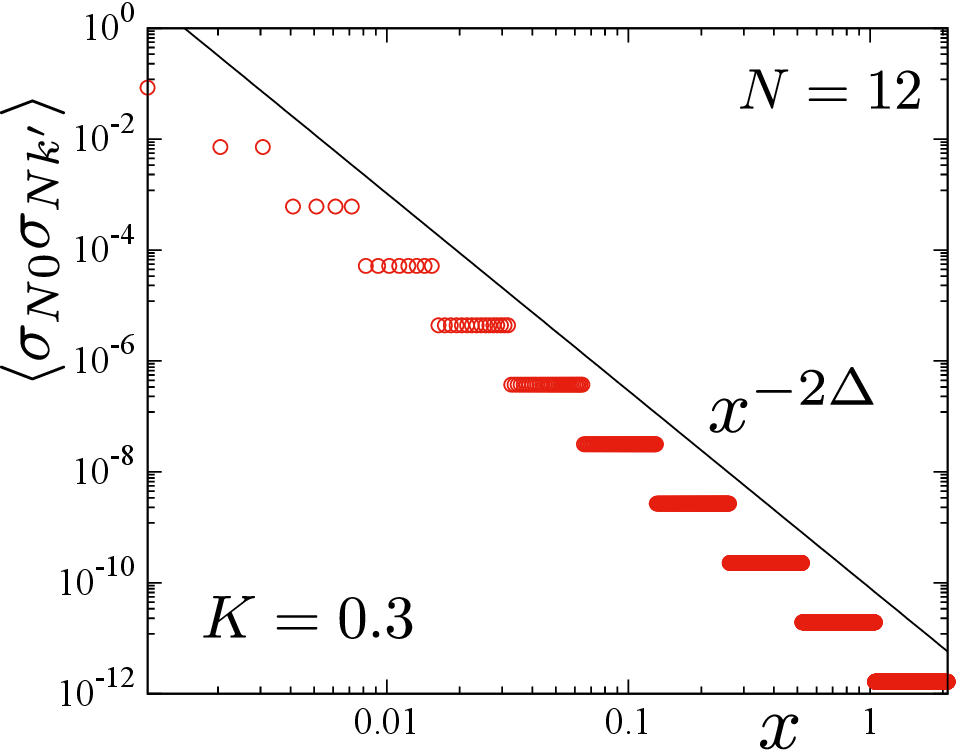}~~
\end{center}
\caption{The correlation function of the Cayley tree Ising model ($q=3$) for $K=0.3$ with $N=12$.
The location of spin pair $(k=0,k'=7)$ is specified by the endpoints of the red path in Fig. \ref{fig1}.
The solid line is a guide indicating the slope of $x^{-2\Delta}$ with $K=0.3$.}
\label{fig3}
\end{figure}

Combining Eqs. (\ref{eq_corr}) and (\ref{eq_def_r}), we obtain 
\begin{align}
\langle \sigma_{N0} \sigma_{Nk'} \rangle  \sim \frac{1}{(x_{k'}-x_0)^{2\Delta}} \, ,
\label{eq_corr_r}
\end{align} 
which also shows a power-law decay with respect to $x$.
Figure \ref{fig3} shows a log-log plot of $\langle \sigma_{N0} \sigma_{Nk'}\rangle$ as a function of $x$ for $q=3$ with $N=12$, where $k=0$ is fixed at the rightmost branch while $k'$ runs along the outer edge cricle up to $k'=2^{11}-1$.
As in the case of $\tilde{h}_n$, the exponential decay in Eq. (\ref{eq_corr}) has been converted to the power-law behavior with the scaling dimension $2\Delta$. 
Note that the step-way-like structure reflects the fact that the descendent branches having the same $n_{0k'}$ (and thus the same $d_{0k'}$) always give rise to the same value of the spin correlation.

In accordance with the scaling property of $\tilde{h}_n$, it is also important to examine the holographic RG description of the local spin operator $\sigma_N$, which is directly coupled with the boundary magnetic field $h_b$.
As in the case of $\tilde{h}_n$, we pick up an effective chain connecting $\sigma_0$ and $\sigma_N$ and insert $\sigma_N$ at the edge of the chain.
Taking the partial sum with respect to $\sigma_N$, we then extract how the spin operator is renormalized toward the interior. 
\footnote{If $h_b=0$, the expectation value of $\sigma_N$ becomes exactly zero for $T>T_B$. 
The aim of the analysis here is to extract how $\sigma_N$ operator can be scaled in the holographic RG process.}
Note that the other branches on the lattice do not affect the renormalization of $\sigma_N$.
Using the identity $e^{K\sigma \sigma'} = \cosh K (1+\tanh K \sigma \sigma')$, we explicilty have
\begin{align}
 & \sum_{\sigma_0\cdots \sigma_{N}} \sigma_N  e^{K \sum_{i=0}^{N-1} \sigma_{i}\sigma_{i+1}}   \nonumber \\
 &= 2\cosh K \!\!\! \sum_{\sigma_0\cdots \sigma_{N-1}} \!\! [\tanh K \sigma_{N-1} ]e^{K \sum_{i=0}^{N-2} \sigma_{i}\sigma_{i+1}} \, \nonumber \\
 &\cdots \nonumber \\
 &=  (2\cosh K)^{N-n} \!\!\! \sum_{\sigma_0\cdots \sigma_{n}} [ (\tanh K)^{N-n} \sigma_{n}] e^{K \sum_{i=0}^{N-n+1} \sigma_{i}\sigma_{i+1}}\,
\end{align}
where the overall coefficient $(2\cosh K)^{N-n}$ gives the same contribution to $C_{N-n}$ in the paritition function (\ref{eq_bethe_prt}) with $h_b=0$.
Thus we can read off the equivalence of $\sigma_n$ and $\sigma_N$ as
\begin{align}
(\tanh K)^{N-n} \sigma_{n} = \sigma_N\, ,
\end{align}
at the spin operator level.
From this relation, it follows that the scaling relation for the spin operator at the $n$th generation is 
\begin{align}
\sigma_n = (\tanh K)^{-N+n} \sigma_N \sim z^{\Delta} \sigma_N \, ,
\label{eq_spin_exp}
\end{align}
which is consistent with the correlation function (\ref{eq_corr_r}).

\subsection{connection to the Poincare metric}
\label{Poincare}

Next, we would like to discuss the property of the present disk coordinate $(z_n, x_k)$ in connection with the Poincare coordinate. 

We introduce the metric on the Bethe lattice such that the each lattice point has a unit area. 
As $n$ shifts as $n\to n+1$, the length increases by one. 
Also the length of the circle at a fixed value of the radius $R=1-z$ is $qp^{n-1}$. 
Therefore the metric of our Bethe lattice looks like
\begin{align}
ds^2 = \frac{L^2}{z^2}(dz^2+ d\tilde{x}^2 )\, .
\label{eq_ads2}
\end{align}
where 
\begin{align}
L\equiv \frac{1}{\log p} \, , 
\label{eq_pradi}
\end{align}
and we have rescaled $\tilde{x} = \frac{q\log p}{2\pi p} x $.
In the large $n$ region or in the non-compact limit,  the metric (\ref{eq_ads2}), which is the Poincare half plane metric, correctly describes that of the Poincare disk namely a Euclidean AdS$_2$. 
For $n=0$, we just need to be more careful about the detailed lattice structure and relation between $R_n$ and $z_n$ to reproduce the precise Poincare disk metric.

This suggests that the holographic RG flow of $\tilde{h}_n$ and the correlation function have a certain connection to physics in AdS/CFT.
We further discuss the implication of the Poincare coordinate Eq. (\ref{eq_ads2}) to the Bethe lattice model, focusing on the role of $\tilde{x}$ coordinate.
As mentioned above Eq. (\ref{eq_def_x}), every branch in the Bethe lattice model is equivalent to each other, and thus $x_k$ was just introduced for arranging spin nodes uniformly on the circle with a constant $R_n$. 
In other words, the $\tilde{x}$ coordinate in Eq. (\ref{eq_ads2}) can correctly capture the relation of the leftmost and rightmost branches for a particular subtree network, like the red-line path in Fig. \ref{fig1}.
Then, the distance along the network path between the edge spin pair of $k=0$ and $k'=p^{d_{0k'}}$ is described by the geodesic on Eq. (\ref{eq_ads2}).
More precisely, the geodesic line is given by a semicircle through $\tilde{x}=0$ and $\tilde{x}_{k'}$, and its length is calculated to be 
\begin{align}
\gamma  =  L \int  \sqrt{1+\frac{d\tilde{x}}{dz}^2}\frac{dz}{z} = 2 L \log\left( \frac{\tilde{x}_{k'}}{\epsilon}\right)
\label{eq_geo}
\end{align}
where $\tilde{x}_{k'}=\frac{p^{d_{0k'}}\log p }{p^N } $.
Here, note that $\epsilon$ is a cutoff scale associated with the $z=0$ limit, which is well defined as $\epsilon = z_N = p^{-N}$ for the Bethe lattice model.
Thus, we can reproduce the length of the correlation path at the lattice level  
\begin{align}
\gamma = 2d_{0k'} 
\, ,
\label{eq_gamma_d}
\end{align}
which is consistent with Eq. (\ref{eq_corr}) and Eq. (\ref{eq_def_d}) with $k=0$ and $k'= p^{d_{0k'}}$ .
On the other hand, we should also remark that Eq. (\ref{eq_gamma_d}) does not hold for branches other than the above particular situation because the tree type network breaks the translational invariance in the circumference direction.

Here, one may wonder about the role played by our tree tensor network in its interpretation as a version of AdS/CFT. 
In the standard tensor network interpretation of AdS/CFT, such as the MERA-type network \cite{MERA2007,TNR2015}, a vacuum state in the dual CFT is realized by a given tensor network, which produces the logarithmic correction of area law of entanglement entropy characteristic for two-dimensional CFTs in the case of AdS$_3/$CFT$_2$.
However, the Bethe lattice model, which we consider in this paper, is a typical example of the tree tensor network, where a system part in the outer edge boundary is connected to its complement through only the single bond with the bond dimension $\chi=2$.\cite{TTN2006,Hikihara2023,EBP2023} 
Such a tree tensor network does not show the logarithmic violation of area law, and its quantum state does not seem to correspond to a CFT vacuum. 
Nevertheless, the correlation function (for the particular spin pair at $k=0$ and $k'= p^{d_{0k'}}$), after a certain smearing procedure, interestingly shows the power law decay like a CFT and is governed by the correlation path defined by Eq. (\ref{eq_geo}). As we will argue later, this system is more properly described by a p-Adic CFT. 
It is more direct to interpret our spin system on the Bethe tree as a specially discretized version of a scalar field on AdS$_2$. This is a scalar field on a Bethe tree geometry and defines the bulk side of the p-Adic AdS/CFT.\cite{Gubser2017,Heydeman:2016ldy}

\section{The Bethe lattice Ising model v.s. a scalar field in AdS$_2$ }

On the basis of the scaling property extracted from the Bethe lattice Ising model, we next discuss its correspondence to a scalar field in AdS.
We also discuss a more direct relationship with the p-adic AdS/CFT through the Burhat-Tits tree.

\subsection{Comparison with the scalar filed in AdS$_2$}
Let us consider a scalar field $\phi$ of a mass $m$ in the AdS$_{d+1}$, where $d$ is the dimension of a boundary CFT.
We then compare  Eqs. (\ref{eq_h_z}), (\ref{eq_Delta}) and (\ref{eq_corr_r}) with a scalar field in the AdS$_{d+1}$, whose scaling property  near the AdS boundary($z \to 0$) is described by
\begin{equation}
\phi(z) \sim A z^{\Delta_-} + B z^{\Delta_+} \, , 
\label{eq_adscft}
\end{equation}
with
\begin{align}
\Delta_\pm \equiv \frac{d}{2} \pm \sqrt{m^2 + \frac{d^2}{4}}\, .
\label{eq_Delta_ads}
\end{align}
In the context of AdS/CFT, $A$ represents a boundary field coupled to the boundary CFT, and $B$ gives a response of the operator against a perturbation with respect to $A$.
Also, note that   
\begin{align}
\Delta_+ + \Delta_- =d
\label{eq_delta_d}
\end{align}
always holds for the scalar field.

Now we compare the results for the Bethe-lattice Ising model with Eq. (\ref{eq_adscft}). 
First, $h_b$ and its conjugate operator $\sigma_N$ correspond to the source and response fields for the scalar field. 
Next, Eq. (\ref{eq_delta_d}) with $d=1$ always holds for the Bethe lattice system.
This behavior is attributed to the geometrical reason that the branching number $p$ and the scaling of $z_n$ are balanced in the disk representation.
Third, $d=1$ is consistent with that the Bethe lattice system with no quantum fluctuation is mapped onto an asymptotic Poincare coordinate. 
The correspondence can be summarized as
\begin{align}
 d & = 1 \,,  \\
 A & \leftrightarrow h_b \,, \\
 B & \leftrightarrow \sigma_N  \,, \\
 \Delta_+  &\leftrightarrow \Delta \, .
\end{align}
In addition, we should remark that the Ising coupling $K$ is invariant in the renormalization process of $\tilde{h}_n$ and $\sigma_n$.
This implies that $K$ can be regarded as a controlling parameter of thermal fluctuations behind $\tilde{h}_n$ and $\sigma_n$, like a mass $m$ for the scalar field.

\begin{figure}[tb]
\begin{center}
\includegraphics[width=6.5cm]{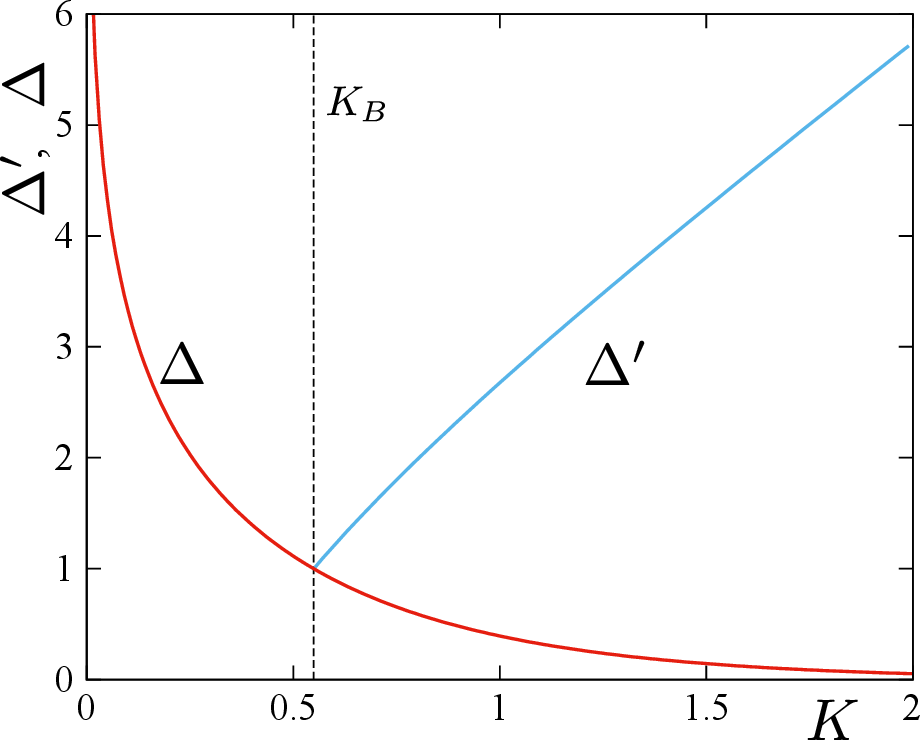}~~
\end{center}
\caption{The scaling dimensions $\Delta$ and $\Delta'$ for the Cayley tree Ising model($q=3$).
The vertical broken line indicates the Bethe transition point $K_B=\arctan 1/2 = 0.549\cdots$ 
}
\label{fig4}
\end{figure}

We analyze properties of the scaling dimension $\Delta$ of Eq. (\ref{eq_Delta}) in detail.
Figure \ref{fig4} shows the $K$-dependence of $\Delta$ for the Cayley tree model, where $\Delta$ runs from zero to infinity as $K$ increases.   
We can then verify $\Delta=1$ at the Bethe transition point $K_B= \arctan 1/2 = 0.549\cdots$ with the use of Eq. (\ref{eq_Tb}). 
For the scalar field, $\Delta$=1 corresponds to the massless point $m=0$.
Clearly, $\tilde{h}$ behaves as a marginal field at $T_B$, which is consistent with the massless scalar field case.
Expanding $\Delta$ around $K=K_B$, moreover, we have 
\begin{align}
\Delta \simeq 1 - \kappa_p (K-K_B) +\mathcal{O}((K-K_B)^2)\, , 
\label{eq_Delta_KB}
\end{align}
where $\kappa_p \equiv \frac{(p^2-1)}{p\log p} $.
For $|m^2| \ll 1 $, then, we can identify  
\begin{align}
m^2 \equiv \kappa_p(K_B-K)\, ,
\label{eq_mass_eff}
\end{align}
or $m \propto |K_B-K|^{1/2}$, which is also consistent with the mean-field universality of the Bethe approximation.

In the high-temperature region ($K<K_B$),  $\tilde{h}_*=0$ is a stable fixed point of the interior,  where $\tilde{h}_n$ behaves as an irrelevant field with increasing $z$. 
This behavior corresponds to the positive $m^2$ in Eq. (\ref{eq_mass_eff}).
For $K>K_B$, on the other hand, the negative $m^2$ suggests a tachyon mode, reflecting that $\tilde{h}_*=0$ is an unstable fixed point and $\tilde{h}_n$ flows to the nontrivial fixed point corresponding to the magnetic order in the sense of the Bethe approximation [Fig. \ref{fig2}(b)].
Then,  Eq. (\ref{eq_Delta}) runs over real-positive values for $0<K<\infty$, which is consistent with the unitarity bound for the scalar field case.

Here we would like to emphasize that the sum of the exponents in Eqs. (\ref{eq_h_z}) and (\ref{eq_spin_exp}) always satisfy Eq. (\ref{eq_delta_d}) with $d=1$ away from the massless point $K=K_B$.
This property reflects that the relationship of Eq. (\ref{eq_delta_d}) originates from the geometrical reason, i.e., the connectivity of the Bethe lattice network independent of details of the Hamiltonian for physical spin degrees of freedom.

\subsection{Relation to the p-Adic AdS/CFT}
\label{subsec_padic}

To be exact, the two point function (\ref{eq_corr_r}) in the Bethe lattice model takes the form of the power law decay as in the CFT only when $x_k$s are situated at special locations of lattice points. 
To obtain the genuine power law we need to smear the $x$ coordinate. 
Instead, the holographic relation on the Bethe lattice can be interpreted as the p-adic AdS/CFT \cite{Gubser2017,Heydeman:2016ldy} more precisely as we will explain below.

By choosing a prime number $p$, the p-adic numbers, expressed as $\mathbb{Q}_p$ are defined by a completion of the rational number $\mathbb{Q}$ with respect to the p-adic norm $|*|_p$. \cite{Brekke1993}
The p-adic norm is defined such that when $y$ is written as $y = a/b \cdot p^m $, we have $|y|_p = p^{- m}$. 
Here $a$ and $b$ are coprime integers, each of which is not divisible by $p$. 
In particular we have $|p^k|_p = p^{-k}$ for any positive integer $k$. 

First, we note that the boundary correlation function in the Ising model on the Bethe lattice agrees with that in the p-adic CFT:
\begin{align}
\langle \sigma_y \sigma_{y'}\rangle \sim
\frac{1}{|y - y'|_p^{2\Delta} } 
\end{align}
with $p(=q-1)$, where $\Delta$ is the conformal dimension of $\sigma$.
\footnote{Note that $y(=$ integer) corresponds to $k$ in Fig. \ref{fig1} rather than $x_k$ in Eq. (\ref{eq_def_x}).}  
Moreover, we can interpret the magnetic field $h$ on the Bethe lattice as the bulk scalar field $\phi_p$ on the p-adic AdS$_2$, i.e. the Burhat-Tits tree shown in Appendix \ref{appendixB}.
The bulk field in p-adic AdS$_{d+1}$ behaves as follows near the AdS boundary $|z|_p \to 0 $:
\begin{align}
\phi_p(z) \sim  |z|_p^{d - \Delta} \, .
\end{align}
Indeed, if we set $z = p^n$,\footnote{Here, $z=p^n$ indicates a digit of a p-adic number, which is distinct from $z_n$ of Eq. (\ref{eq_def_z}) for the disk representation.}
we obtain
\begin{align}
\phi_p(z) \sim  p^{-(d - \Delta)n} \, ,
\end{align}
which agrees with the behavior $\tilde{h}_n \simeq  \lambda^{N - n}h_b$. 
This allows us to identify
\begin{align}
\lambda  = p^{1 - \Delta}\, ,
\end{align}
which agrees with the result (\ref{eq_Delta}) in  the Bethe lattice Ising model.

As in Appendix \ref{appendixB}, we can also directly relate the p-adic norm between two points $y$ and $y'$ at the boundary with the length $d_{yy'}$ in the tree graph via
\begin{align}
|y -y'|_p = p^{-(N-d_{yy'})}\, ,
\label{eq_padic_dist}
\end{align}
where we defined $d_{yy'}$ by the integer which counts how many layers we need to go up in order to connect $y$ with $y'$. 
This is manifest in Fig.\ref{fig8}.
If we adjust the overall normalization by $p^{-N}$,  Eq. (\ref{eq_padic_dist}) is consistent with Eq. (\ref{eq_def_r}) except for the ordering of the node index irrelevant to the intrinsic physics.

\section{Ferromagnetic fixed point and crossorver}

\subsection{Ferromagnetic fixed point}

For $K>K_B$, the tachyon mode is unstable, and $\tilde{h}_n\ne 0$ flows to the nontrivial fixed point of Eq. (\ref{eq_bethe_fixed}), which corresponds to the ferromagnetic order in the deep interior.
We next discuss the property of the stable fixed point for $K>K_B$. 
For this purpose, let us begin with the fact that if $h_b=\tilde{h}_*$, the effective magnetic field $\tilde{h}_n=\tilde{h}_*$ is also homogeneous in the Bethe lattice.
Then we can analyze the scaling property of the fluctuation field $\delta\tilde{h}_n$ defined by $\tilde{h}_n = \tilde{h}_* + \delta \tilde{h}_n$ with $ \delta \tilde{h}_N = h_b - \tilde{h}_* $.
Linearizing the RG relation of Eq. (\ref{eq_betheRG}) around $\tilde{h}_*$, we obtain 
\begin{align}
\delta \tilde{h}_{n} = \delta \tilde{h}_N  \lambda'^{N-n} \, ,
\end{align}
with $\lambda' \equiv f'(\tilde{h}_*)$.
In terms of $z_n$ coordinate, the scaling dimension for $\delta \tilde{h}_{n}$ can be represented as 
\begin{align}
\delta\tilde{h}_n \sim z_n^{1-\Delta'}\, ,
\label{eq_h_z_prime}
\end{align}
with
\begin{align}
 \Delta' = -\frac{\log\left( \frac{\sinh(2K)}{\cosh(2K)+\cosh(2\tilde{h}_*)} \right) }{\log p}\, .
\label{eq_delta_prime}
\end{align}
In Fig. {\ref{fig4}, we plot the $K$ dependence of $\Delta'$ by solving the fixed point equation of (\ref{eq_bethe_fixed}) numerically.
In the figure,  $\Delta'=1$ is confirmed at the transition point $K=K_B$.
For $K>K_B$, moreover, we find $\Delta'> 1 $, implying that $\tilde{h}^*$ is a stable fixed point.

For $h_b=\tilde{h}_*$, the long-distance behavior of the correlation function can also be calculated exactly by the transfer matrix technique\cite{Hu1998}.
Using the effective transfer matrix along the correlation path (See Appendix A for details), we then obtain the exact expression of the correlation function along the outer edge boundary for general $K(>K_B)$ as
\begin{align}
 \langle \sigma_{N0}\sigma_{Nk'} \rangle -  M_*^{2} = (1-M_*^2) x^{-2\Delta'} \, ,
 \label{eq_corr_order}
\end{align}
where $M^*$ denotes the expectation value of the ferromagnetic order parameter.
This behavior is consistent with Eq. (\ref{eq_corr_r}).
Thus the relationship of Eq. (\ref{eq_adscft}) can be also confirmed for the fluctuation fields around the ferromagnetically ordered background.

We investigate the critical behavior of the fixed point near $K_B$, where we can explicitly construct the analytic solution of Eq. (\ref{eq_bethe_fixed});
Assuming  $ K-K_B \ll 1$  and thus $\tilde{h}_* \ll 1$,  we expand Eq. (\ref{eq_bethe_fixed}) up to $\tilde{h}_*^{3}$,
\begin{align}
\tilde{h}_*= p \tanh(K) (\tilde{h}_* - \frac{1}{3\cosh(K)^2}  \tilde{h}_*^{3}) \, .
\end{align} 
and obtain
\begin{align}
\tilde{h}_* = \pm \sqrt{3p(K-K_B)} \, ,
\label{eq_h_nearKB}
\end{align}
which corresponds to the mean-field universality of the Bethe approximation.
Then, assuming $\delta \tilde{h}_N \ll  K-K_B$ further, we obtain
\begin{align}
\lambda' = 1 -\frac{2(p^2-1)}{p}(K-K_B)\, .
\end{align}
From this, it follows that $\delta \tilde{h} \propto z_n^{1-\Delta'}$ with the scaling dimension
\begin{align}
\Delta' = 1 +  2\kappa_p(K-K_B) \, ,
\label{eq_Delta'}
\end{align}
which can be also derived by directly expanding Eq. (\ref{eq_generalDelta'}) with respect to $K-K_B$ togeather with Eq. (\ref{eq_h_nearKB}). 
An important point is that the sign in front of the $K-K_B$ term is inverted from Eq. (\ref{eq_Delta_KB}) and thus $\Delta'>1$ for $K>K_B$, which is consistent with Fig. \ref{fig4}.
Note the coefficient $2\kappa_p$ in Eq. (\ref{eq_Delta'}) represents the slope of $\Delta'$ in the vicinity of $K_B$ in the figure.

\subsection{Crossorver}
\label{subsec_crossover}

\begin{figure}[tb]
\begin{center}
\includegraphics[width=6.5cm]{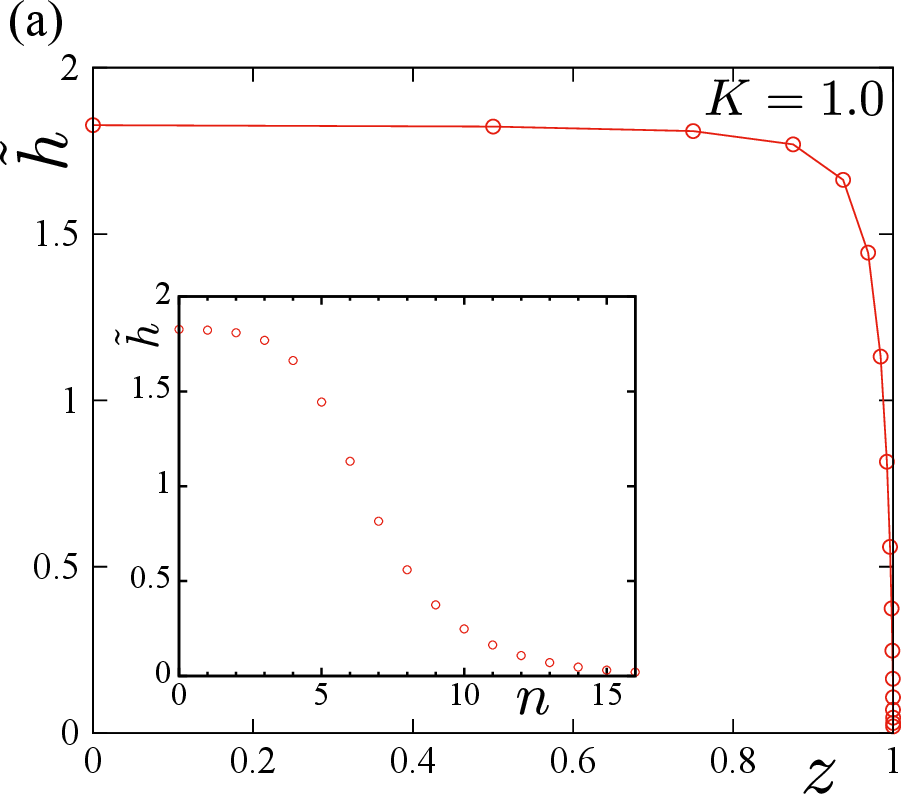}
~~~~~
\includegraphics[width=5.5cm]{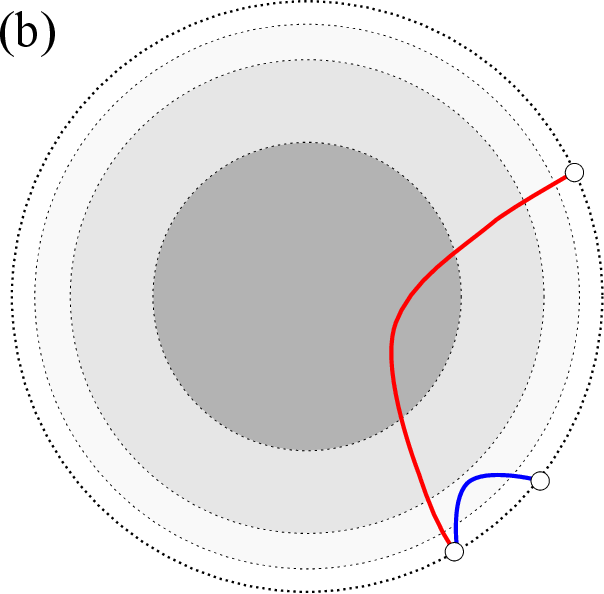}
\end{center}
\caption{(a) $z$ dependence of the effective magnetic field $\tilde{h}$ for $K=1.0$ with $N=16$.
Inset: $n$-dependence of $\tilde{h}$, where $h_b=0.02$ corresponds to the boundary magnetic field. 
(b) Schematic diagram for a crossover behavior of a correlation path.}
\label{fig5}
\end{figure}

\begin{figure}[tb]
\begin{center}
\includegraphics[width=7cm]{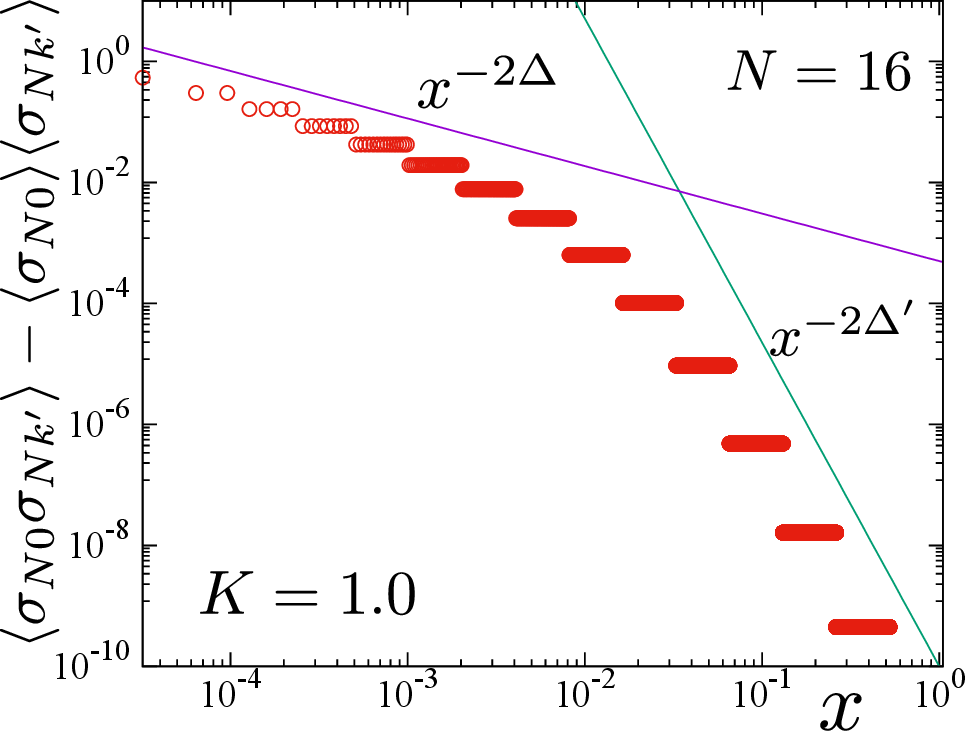}~~
\end{center}
\caption{The correlation function of the Cayley tree Ising model ($q=3$) for $N=16$ with $K=1.0$.
$\langle \sigma_{N0} \sigma_{Nk'} \rangle - \langle \sigma_{N0} \rangle \langle \sigma_{Nk'} \rangle$. 
The location of spins is specified by the endpoints of the red path in Fig. \ref{fig1}.
The solid lines are a guide for eyes indicating the slope of $x^{-2\Delta}$ and $x^{-2\Delta'}$ for $K=1.0$.
$\langle \sigma_{N0} \rangle = \langle \sigma_{Nk'} \rangle = 0.2069\cdots$.   }
\label{fig6}
\end{figure}

We next turn to the case where $h_b$ is small but finite.
In this case, $\tilde{h}(\ll \tilde{h}_*)$ crossovers from the unstable tachyon mode to $\tilde{h}_*$ in the deep interior, as $z_n$ increases[See Fig. \ref{fig2}(b)].
We actually compute a series of $\tilde{h}_n$ starting from $h_b=0.02$ for $K=1.0$ and $N=16$ and show the resulting flow with respect to $z$ in Fig. \ref{fig5}(a).
As depicted in Fig. \ref{fig5}(b), then, we can expect that the correlation function also crossover from Eq. (\ref{eq_corr_r}) to Eq. (\ref{eq_corr_order});
As $x$ increases, the correlation path enters the deep interior where the spin has a stable magnetic order $M_*$.
We perform a numerical computation of the correlation function for the Cayley tree model($q=3$) of $N=16$.
Figure \ref{fig6} shows the numerical result for $h_b=0.02$ and $K=1.0$, for which the expectation value of the edge spin is obtained as $\langle \sigma_{N0}\rangle = \langle \sigma_{Nk'} \rangle = 0.2069\cdots$.
The computational details are described in Appendix \ref{appendix_a}.

As in Fig. \ref{fig6}, the correlation function exhibits a clear crossover behavior.
In the inset of Fig. \ref{fig5}(a), one can find $\tilde{h}_n <0.2$ ($d_{0k'}=N-n_{0,k}< 5$) for $n>10$, which governs a short distance behavior in $x < 0.003$, 
Then, the power-low decay of the correlation function is clearly characterized by $2\Delta$ associated with the tachyon mode around the $\tilde{h}_*=0$ fixed point. 
This situation corresponds to the blue correlation path in Fig. \ref{fig5}(b).
As $x$ increases, the correlation path penetrates the deep interior, like the red line in Fig. \ref{fig5}(b), and thus the long-distance behavior is governed by the ferromagnetic fixed point of $\tilde{h}_n = \tilde{h}_* = 1.824\cdots $.
Then the correlation function is characterized by the fluctuation around $\tilde{h}_*$ with the power $2\Delta'$, which can be verified in Fig. \ref{fig6}.

Here, we note that the above behavior of the correlation function is a crossover rather than a phase transition since the bulk free energy of the Bethe lattice Ising model has no singularity at a finite temperature, although the fixed point solution of Eq. (\ref{eq_bethe_fixed}) [the self-consistent solution in the Bethe approximation terminology] shows the mean-field type phase transition. 
In the language of the tree tensor network, the above property of the Bethe lattice model can be attributed to the fact that the position of root (top) tensors has no intrinsic role in the tree tensor network because it can be shifted to any link by singular decomposition without loss of generality.\cite{TTN2006,EBP2023,Hikihara2023}
In fact, the RG transformation for $\tilde{h}$ may not be in the direction from the boundary (ultraviolet) to the interior (infrared). 
\footnote{For example, this is the case for $\tilde{h}'_1$ with a red arrow in Fig. \ref{fig7}.}
In particular, if $\tilde{h}$ is at the fixed point, $\tilde{h}^*$ becomes uniform on the entire network independently of the direction of the RG transformation.
Thus, the correlation functions exhibit the crossover behavior even if the phase transition in the sense of the Bethe approximation exits in the deep interior.

 \section{conclusion and outlook}
 
We have elucidated the holographic aspect of the Bethe lattice Ising model, which provides a simple statistical-mechanics description of the holographic RG in AdS$_2$.
In particular, the 1D nature of the Bethe lattice model enabled us to analytically formulate the RG transformation for the effective magnetic field $\tilde{h}_n$ from the outer boundary to the interior and to calculate the boundary spin correlation functions exactly.
By introducing the Poincare disk-like coordinate, we clarified the underlying mechanism of the power-law decay after a suitable smearing and the cutoff scale for the Bethe lattice network to extract the exact scaling dimensions.
Then, we confirmed the consistency of the Bethe lattice model with the AdS$_2$/CFT$_1$.
More precisely, we clarified that this Bethe lattice model is directly interpreted as the p-adic AdS/CFT\cite{Gubser2017}, which may provide an intriguing toy model of AdS/CFT through its tree tensor network representation\cite{Heydeman:2016ldy, Bhattacharyya:2017aly}.
Moreover, it was demonstrated that the phase transition in the interior induces a nontrivial crossover of boundary spin correlation functions, depending on the depth of the corresponding correlation path.
Nevertheless, we should also note the CFT side corresponding to the Bethe lattice model remains a highly nontrivial problem, as mentioned in Sec. \ref{Poincare}.
In terms of the tensor network, a nonuniversal transitional layer may be required to bridge the scale-invariant bulk network with the spin model at the boundary\cite{Evenbly2013Book}.

Our study has presented a simple theoretical structure realizing the holographic RG from the statistical-mechanics side.
At the same time, we think this novel standpoint opens several possible research directions toward a deeper understanding of the tensor network description of the AdS/CFT.
For instance, the present work has an essential return to the dimensionality of the Bethe lattice model. 
At the level of the Bethe approximation, the coordination number $q$ corresponds to the spatial dimension of the regular lattice.
However, the resulting phase transition for the interior of the Bethe lattice model is of the mean-field universality class independently of $q$, and thus the coordination number loses its original meaning.
On the other hand, the spatial dimension $d$ in the context of AdS$_{d+1}$/CFT$_d$ can be controlled by the ratio of the branching number $p$ in Eq. (\ref{eq_lambda}) and the scaling factor (\ref{eq_def_z}) for the effective disk coordinate.
By adjusting $p$ to $p^d$ in Eq. (\ref{eq_lambda}), one can actually modify the spatial dimension of the effective coordinate from $1$ to $d$, as also pointed out for the p-adic AdS/CFT\cite{Gubser2017,Heydeman:2016ldy}.
Such an interplay of two kinds of dimensionalities between the interior and the boundary states may be a new insight for designing tensor networks in higher dimensional systems.

Another important research direction is the tensor network with the loop network structure, which is essential to reproduce the area-law of entanglement entropy up to the log correction, as in the case of MERA\cite{MERA2007,Swingle2012}. 
In particular, the hyperbolic lattice model, which simultaneously realizes a regular loop structure and a translational symmetry, is a fundamental target beyond the Bethe-type tree network, where interactions inserted among tree branches would unfold such a degenerating structure of the correlation functions observed in Figs. \ref{fig3} and \ref{fig6}.
Indeed, several numerical studies have been performed for various hyperbolic tiling models.\cite{Shima2006,UedaK2007,Krcmar2008,Serina2016} 
Recently,  moreover, the numerical evidence of nontrivial scaling dimensions for boundary-spin correlation functions was obtained by Monte Carlo simulations.\cite{Asaduzzaman2020}
Similarly, matchgate tensor networks based on the specific tensors exhibited the power-law decay of boundary spin correlators.\cite{Jahn2019}
In addition, a holographic RG argument based on an inflation rule for hyperbolic tilings suggests a quasi-periodic spin chain as a corresponding boundary theory.\cite{Basteiro2022}
For analyzing holographic aspects of hyperbolic tiling models quantitatively, the corner-transfer-matrix renormalization group\cite{CTMRG1,UedaK2007} may be a suitable tensor network method enabling us to calculate the boundary spin correlations and the entanglement structure precisely.\cite{ON2023}
In turn, by adding the transverse field term to bulk nodes in the Bethe lattice network, one can also investigate the role of the (imaginary) time coordinate that is missing at the level of the classical Ising model.
This step would be essential to improve our understanding of the holographic RG into AdS$_3$.
However, the exact RG transformation, like the classical Ising model, is generally hard, where such a tensor network method as density matrix renormalization group that can directly treat quantum spin systems on the Bethe lattice\cite{Otsuka1996} would be essential.

Recently, a Bethe-lattice or equivalently Burhat-Tits-tree-network system was actually constructed in a setup of ultra-cold-atomic-gas experiment.\cite{Bentsen2019}
 Also,  a circuit quantum electrodynamics experiment realized a hyperbolic lattice system.\cite{Kollar2019}
Possibly, these developments suggest that holographic tensor networks could be tested in a realistic experimental situation near future.
This work on the holographic RG would mediate further studies of such systems at the nexus of quantum gravity and condensed matter experiments.


\section*{acknowledgments}
The authors thank T. Nishino for valuable discussion.
This work is supported by Grants-in-Aid for Transformative Research Areas "The Natural Laws of Extreme Universe---A New Paradigm for Spacetime and Matter from Quantum Information" (KAKENHI Grant No. JP21H05182, No. JP21H05187 and No. JP21H05191) from MEXT of Japan.
It is also supported by Inamori Research Institute for Science and by Grant-in-Aid for Scientific Research (KAKENHI Grant No.JP21H04469) from JSPS.


\appendix

\section{Correlation functions}
\label{appendix_a}

For the Bethe lattice Ising model, the calculation of the correlation function can be reduced to that for an effective 1D Ising chain. 
As an example, let us consider the correlation path (red line) in Fig. \ref{fig1}, which contains $2d_{kk'}+1$ number of spins and thus $2d_{kk'}$ number of bonds with $d_{kk'}=3$.
Then, the contribition from the subtree branches attached to the correlation path are always renormalized into the effective magnetic fields for spins along the chain. 
This situation is actually illustrated in Fig. \ref {fig7}, which shows the 1D Ising chain in the fields $\{\tilde{h}_n \}$ along the correlation path.
Here, note that $\tilde{h}'_1$, which corresponds to the top node in the correlation path,  comes from the upper branch of the network, but it can be also evaluated by using the recursive relation of Eq. (\ref{eq_betheRG}).

We then represent the partition function with the use of the transfer matrix technique,
\begin{align}
{\cal Z}_{N} = \vec{u}^\dagger T^\dagger_{1} \cdots T^\dagger_{d_{kk'}} T^{~}_{d_{kk'}} \cdots T^{~}_{1} \vec{u}\, .
\label{eq_eff1d}
\end{align} 
The transfer matrix is explicitly defined by
\begin{widetext}
\begin{align}
T_{i} = 
\begin{pmatrix}
e^{K+\frac{(p-1)}{2p}(\tilde{h}_{N-i} + \tilde{h}_{N-i+1})} & e^{-K+\frac{p-1}{2p}(\tilde{h}_{N-i} - \tilde{h}_{N-i+1})} \\
e^{-K-\frac{p-1}{2p}(\tilde{h}_{N-i} - \tilde{h}_{N-i+1})} & e^{K-\frac{p-1}{2p)}(\tilde{h}_{N-i} + \tilde{h}_{N-i+1})}
\end{pmatrix} 
\label{eq_def_tm}
\, ,
\end{align}
for $i\ne d_{kk'}$, and 
\begin{align}
T_{d_{kk'}} = 
\begin{pmatrix}
e^{K+\frac{(p-1)}{2p}(\tilde{h}'_{N-d_{kk'}} + \tilde{h}_{N-d_{kk'}+1})} & e^{-K+\frac{p-1}{2p}(\tilde{h}'_{N-d_{kk'}} - \tilde{h}_{N-d_{kk'}+1})} \\
e^{-K-\frac{p-1}{2p}(\tilde{h}'_{N-d_{kk'}} - \tilde{h}_{N-d_{kk'}+1})} & e^{K-\frac{p-1}{2p}(\tilde{h}'_{N-d_{kk'}} + \tilde{h}_{N-d_{kk'}+1})}
\end{pmatrix} 
\,  ,
\label{eq_def_tmtop} 
\end{align}
for $i=d_{kk'}$.
Here, $\tilde{h}'_{N-d_{kk'}}$ denotes the contribution of the effective field from the upper branches.
\end{widetext}

In Eq. (\ref{eq_eff1d}), we have also introduced the boundary vector,
\begin{align}
\vec{u} = 
\begin{pmatrix}
e^{\frac{p+1}{2p}\tilde{h}_{b}} \\
 e^{-\frac{p+1}{2p}\tilde{h}_{b}}
\end{pmatrix}\, ,
\label{eq_boundary_u}
\end{align}
to adjust the boundary magnetic field.
Here note that no branch is attached to the boundary spin.

In the framework of the transfer matrix, the expectation value of the edge spin and the edge spin correlation function
are respectively calculated as
\begin{align}
\langle \sigma_{Nk'}  \rangle &= \mathcal{Z}'/ \mathcal{Z}_{N} \\
\langle \sigma_{Nk} \sigma_{Nk'} \rangle &= \mathcal{Z}''/ \mathcal{Z}_{N}
\label{eq_eff1d_corr}
\end{align}
where
\begin{align}
\mathcal{Z}'\equiv \vec{u}^\dagger \sigma^z T_1^\dagger \cdots  T^\dagger_{d_{kk'}} T^{~}_{d_{kk'}} \cdots {T}^{~}_1   \vec{u} \, ,
\end{align}
and
\begin{align}
\mathcal{Z}''\equiv \vec{u}^\dagger \sigma^z T^\dagger_1 \cdots  T^\dagger_{d_{kk'}} T^{~}_{d_{kk'}} \cdots {T}^{~}_1  \sigma^z \vec{u} \, .
\end{align}
Here, $\sigma^z$, the $z$ component of the usual Pauli matrix, are formally inserted corresponding to the Ising spins $\sigma_{Nk}$ and $\sigma_{Nk'}$ at the edges.

\begin{figure}[bt]
\begin{center}
\includegraphics[width=7cm]{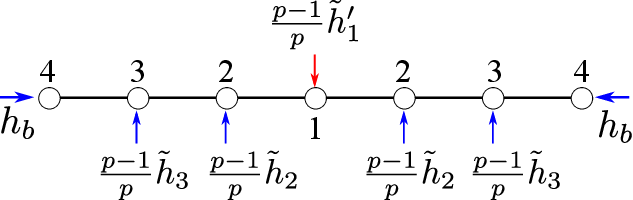}~~
\end{center}
\caption{The effective 1D chain structure for the correlation path in Fig. \ref{fig1}.
The number at each spin indicates the generation index counting from the center site. 
The arrows represent the renormalized effective magnetic field from the branches attached at each site.
In particular, the red arrow for $\tilde{h}'_{1}$ comes from the upper branches.}
\label{fig7}
\end{figure}

\subsection{$h_b=0$} 
If $h_b=0$, the renormalized magnetic field $\tilde{h}_n=0$ everywhere in the Bethe lattice, implying that the effective chain has no magnetic field. 
Then, Eq. (\ref{eq_eff1d_corr}) is straightforwardly obtained as
\begin{align}
\langle \sigma_{Nk}\sigma_{Nk'}\rangle = (\tanh K)^{2d_{kk'} } \, ,
\end{align}
which corresponds to Eq. (\ref{eq_corr}) in the main text.

\subsection{$h_b=\tilde{h}_*$}

For $h_b=\tilde{h}_{*}$, $ \tilde{h}_n = \tilde{h}_*$ is also satisfied for all $n$ in the effective chain, implying that $T$-matrix in Eq. (\ref{eq_eff1d}) is symmetric and homogeneous.
Then the correlation function for Eq. (\ref{eq_eff1d}) can be analytically calculated with the diagonalization of 
\begin{align}
T_* = 
\begin{pmatrix}
e^{K+\frac{p-1}{p}\tilde{h}_* } & e^{-K} \\
e^{-K} & e^{K-\frac{p-1}{p}\tilde{h}_* } 
\end{pmatrix} \, .
\end{align}
$T_*$ is formally equivalent to the usual 1D Ising chain in a magnetic field.
However, the effective magnetic field $\tilde{h}_*$ satisfies the fixed point condition (\ref{eq_bethe_fixed}), 
\begin{align}
e^{\frac{2\tilde{h}_*}{p}} = \frac{e^{2K+2\tilde{h}_{*}} +1 }{e^{2K}+e^{2\tilde{h}_{*}} }\,,
\label{eq_id1}
\end{align}
or equivalently
\begin{align}
e^{2\tilde{h}_*} = \frac{e^{2K+\frac{2\tilde{h}_{*}}{p}}-1 }{e^{2K}-e^{\frac{2\tilde{h}_{*}}{p} } } \, ,
\label{eq_id2}
\end{align}
which play a technically essential role to resolve the eigenvalue problem of $T_*$.
Actually one can straightforwardly prove that 
\begin{align}
 T_* \vec{u}_* = \Lambda_1 \vec{u}_*
\end{align}
where 
\begin{align}
\Lambda_1 & = 2 e^{\frac{\tilde{h}_*}{p}} \cosh(K+\tilde{h}_*) \nonumber \\
 &= \frac{2 \sinh(2K) e^{K+\frac{p-1}{p}\tilde{h}_*}}{e^{2K} - e^{-\frac{2\tilde{h}_*}{p}}}  \, ,
\end{align}
and
\begin{align}
\vec{u}_*=
\begin{pmatrix}
e^{\frac{p+1}{2p}\tilde{h}_*} \\
e^{-\frac{p+1}{2p}\tilde{h}_*}
\end{pmatrix}
 \, .
 \label{eq_boundary_u*}
\end{align}
Note that $\Lambda_1$ corresponds to the largest eigenvalue of $T_*$.
In other words, the self-consistent condition in the Bethe approximation ensures the projection to the largest eigenvalue state of $T_*$ by the boundary state (\ref{eq_boundary_u*}).
Then, we explicitly obtain the partition function as 
\begin{align}
\mathcal{Z} = \vec{u}_*^{\dagger}T_*^{2d_{kk'}} \vec{u}_*^{~} = 2 \Lambda_1^{2d_{kk'}} \cosh(\frac{p+1}{p}\tilde{h}_*) \, .
\label{eq_h*Z}
\end{align}
Similarly, it is also straightforward to obtain
\begin{align}
\mathcal{Z}'&= \vec{u}_*^{\dagger} \sigma^z  T_*^{2d_{kk'}} \vec{u}_*  \nonumber \\
 & = \vec{u}_*^{\dagger}   T_*^{2d_{kk'}} \sigma^z \vec{u}_*  \nonumber \\
 & =  2 \Lambda_1^{2d_{kk'}} \sinh(\frac{p+1}{p}\tilde{h}_*)\, ,
\end{align}
which leads to the expectation value of the magnetization 
\begin{align}
M_* = \tanh(\frac{p+1}{p}\tilde{h}_*) = \tanh(\frac{q}{q-1}\tilde{h}_*) \, .
\label{eq_h*M}
\end{align}
Of course, this is consistent with the bulk magnetization in the deep interior region for $K>K_B$.

For calculating the correlation function, the other eigenvalue of $T_*$, i.e. $\Lambda_2$ is also required. 
By constructing the vector $\vec{v}_*$ orthogonal to $\vec{u}_*$, we explicitly obtain the eigenvector belonging to $\Lambda_2$ as, 
\begin{align}
\vec{v}_* = 
\begin{pmatrix}
-e^{-\frac{p+1}{2p}\tilde{h}_*} \\
e^{\frac{p+1}{2p}\tilde{h}_*}
\end{pmatrix}
 \, .
\end{align}
By applying $T_*$ to $\vec{v}_*$, we can directly confirm
\begin{align}
\Lambda_2 &= 2 e^{\tilde{h}_*} \sinh(K-\frac{\tilde{h}_*}{p}) \nonumber \\
 &= \frac{2\left( \cosh(2K) -\cosh(\frac{2\tilde{h}_*}{p})\right)e^{K+\frac{p-1}{p}\tilde{h}_*}}{e^{2K} - e^{-\frac{2\tilde{h}_*}{p}}}  \, .
\end{align}
Taking accout of the normalization for $\vec{u}_*$ and $\vec{v}_*$, we obtain
\begin{align}
T_* = P \begin{pmatrix}\Lambda_1 & 0 \\ 0 &\Lambda_2 \end{pmatrix} P^\dagger
\label{eq_plpt}
\end{align}
where
\begin{align}
P = \frac{1}{\sqrt{2\cosh \frac{(p+1)\tilde{h}_*}{p}}}
\begin{pmatrix}
e^{\frac{p+1}{2p}\tilde{h}_*} & - e^{-\frac{p+1}{2p}\tilde{h}_*} \\
e^{-\frac{p+1}{2p}\tilde{h}_*} & e^{\frac{p+1}{2p}\tilde{h}_*}
\end{pmatrix}
\, .
\end{align}

Using Eq. (\ref{eq_plpt}), we can straightforwardly calculate
\begin{align}
\mathcal{Z}'' =   \Lambda_1^{2d_{kk'}} \frac{2 \sinh^2(\frac{p+1}{p}\tilde{h}_*)}{\cosh(\frac{p+1}{p}\tilde{h}_*)}
+  \Lambda_2^{2d_{kk'}} \frac{2}{\cosh(\frac{p+1}{p}\tilde{h}_*)} \, .
\end{align}
Also using Eqs. (\ref{eq_h*Z}) and (\ref{eq_h*M}), we finally obtain the exact expression of the correlation function as
\begin{align}
\langle \sigma_{Nk} \sigma_{Nk'} \rangle = M_*^2 + \left(\frac{\Lambda_2}{\Lambda_1}\right)^{2d_{kk'}} ( 1- M_*^2)\, ,
\end{align}
with
\begin{align}
\frac{\Lambda_2}{\Lambda_1} = \frac{\cosh(2K) -\cosh(\frac{2\tilde{h}_*}{p})}{ \sinh(2K)}
\,.
\end{align}
After some algebra, we obtain 
\begin{align}
\Delta'= -\frac{\log\left(\tanh K -\frac{\sinh^2\frac{\tilde{h}_*}{p}}{\sinh K \cosh K}  \right)}{\log p} \, .
\label{eq_generalDelta'}
\end{align}
Using Eq (\ref{eq_id2}), moreover, one can prove that Eq. (\ref{eq_generalDelta'}) is equivalent to  Eq. (\ref{eq_delta_prime}).

\subsection{ $h_b\ne 0$, $\tilde{h}_*$}

For general $h_b$, the effective 1D chain is not homogeneous, where diagonalization of the transfer matrix is not useful.
We numerically calculate position-dependent effective magnetic fields and directly perform the matrix multiplication of $T$s to the boundary vector $\vec{u}$.
This is a numerically easy problem since $T$ is just a $2 \times 2$ matrix.
The result was presented in Fig. \ref{fig6}.

\section{An example of the Burhat-Tits tree}
\label{appendixB}

We explain the correspondence between the Bethe lattice and Burhat-Tits tree for the p-adic AdS$_2$ by using the Cayley tree network $p=q-1 =2$ as a typical example. 
As in Sec. \ref{subsec_padic},  we assume that $p$ is a prime number, which is related to the coordination number $q$ in the Bethe lattice through $p=q-1$.
Then, the Bethe lattice and the Burhat-Tits tree basically provide an equivalent tree network.
For the former case, the network is recursively constructed by branching the tree from the interior to the outer side.
Thus, $p$ may not be restricted to a prime number for the Bathe lattice Ising model by construction.
For the latter case, meanwhile, a bottom-up building of the tree network from the p-adic number at the bottom layer is possible with the help of the p-adic norm.

\begin{figure}[tb]
\begin{center}
\includegraphics[width=7cm]{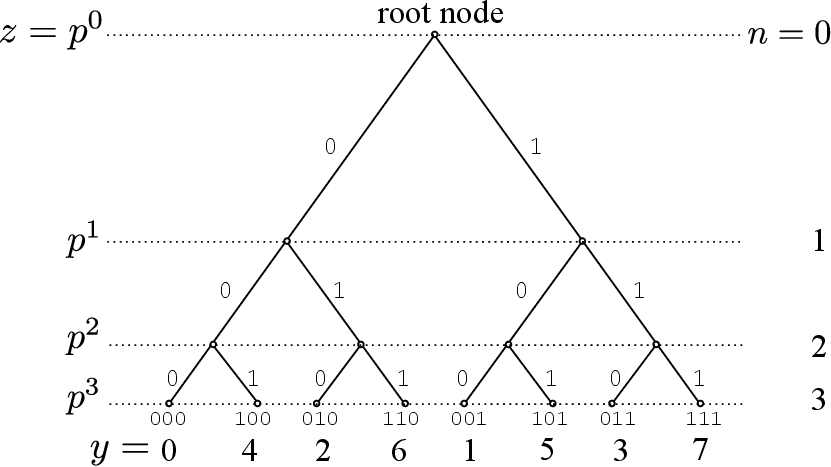}~
\end{center}
\caption{A Burhat-Tits tree for $p(=q-1)=2$ with $N=3$, which is equivalent to the Calylay tree network.
$n$ denotes the index for a digit in a p-adic number, which corresponds to the generation index in the Bethe lattice.
$y$ at the bottom layer denotes the node index constructed from the branching structure of the tree network. 
}
\label{fig8}
\end{figure}

In Fig. \ref{fig8}, we show the Burhat-Tits tree with $p=2$, where $n$ corresponds to the generation index of the Bethe lattice Ising model and the $y$ coordinate at the bottom layer is defined by a p-adic number $y= \sum c_n p^n$, where $c_n = 0, 1, \cdots p-1$.
Note that Fig. \ref{fig8} represents a finite-size network of $N=3$, though an infinite tree is usually treated in the p-adic AdS/CFT.
Then, the distances between $y$ and $y'$ can be explicitly computed in terms of the p-adic norm like
\begin{align}
&|4 - 0|_2 = |6 - 2|_2 = |5 - 1|_2 = |7 - 3|_2 = 2^{-2} \,, \nonumber \\
&|6 - 0|_2 = |6 - 4|_2 = \cdots = 2^{-1}\, , \nonumber \\
&|5 - 0|_2 = |1 - 0|_2 = \cdots = 2^0 \, . \nonumber
\end{align}
This p-adic norm is directly related to the length $d_{yy'}$ in the tree graph via
\begin{align}
|y -y'|_p = p^{-(N-d_{yy'})} (= p^{-n_{yy'}} )\,
\end{align}
where we defined $d_{yy'}$ by the integer which counts how many layers we need to go up in order to connect $y$ with $y'$ as in Eq. (\ref{eq_def_d}).

The ordering of the p-adic variable $y$ in Fig. \ref{fig8} is different from the node index $k$ in Fig. \ref{fig1} along the outer circle of the Bethe lattice. 
However, this difference is not relevant to the physics of the holographic RG in the Bethe lattice Ising model, where every branch is equivalent. 
Actually the $x$ coordinate of Eq. (\ref{eq_def_r}) gives the same distance for the leftmost and rightmost branches of any subtree network in the Bethe lattice, by adjusting the normalization factor $p^{-N}$.

\bibliography{bethe-1}

\end{document}